\documentclass{INTERSPEECH2023}
\usepackage{bm}
\usepackage{amssymb,amsmath,graphicx,bbm,color,booktabs}
\usepackage{amsopn}
\usepackage{caption}
\usepackage{subcaption}
\usepackage{multirow}
\usepackage{xcolor}
\usepackage{amsmath}
\usepackage{hyperref}
\newcommand{\audiotok}{\textsc{AudioToken}}
\usepackage{url}

\renewcommand{\[}{\begin{eqnarray}}
\renewcommand{\]}{\end{eqnarray}}

\newcommand{\R}{\mathbb{R}}

\renewcommand{\eqref}[1]{Eq.~(\ref{#1})}

\interspeechcameraready 

\title{\audiotok: Adaptation of Text-Conditioned Diffusion Models \\ for Audio-to-Image Generation}

\DeclareMathOperator{\E}{\mathbb{E}}
\newcommand\normx[1]{\lVert#1\rVert}

\name{Guy Yariv $^{\heartsuit, \clubsuit}$, Itai Gat $^\diamondsuit$, Lior Wolf $^\spadesuit$, Yossi Adi $^{\heartsuit, *}$, Idan Schwartz $^{\spadesuit, \clubsuit, *}$\thanks{*Equal Contribution.}}

\address{
  $^\heartsuit$The Hebrew University of Jerusalem , $^\diamondsuit$Technion - Israel Institute of Technology  \\ $^\spadesuit$Tel-Aviv University,  $^\clubsuit$NetApp}

\usepackage{hyperref}
\begin{document}
\maketitle

\begin{abstract}
    In recent years, image generation has shown a great leap in performance, where diffusion models play a central role. Although generating high-quality images, such models are mainly conditioned on textual descriptions. This begs the question: \emph{how can we adopt such models to be conditioned on other modalities?}. In this paper, we propose a novel method utilizing latent diffusion models trained for text-to-image-generation to generate images conditioned on audio recordings. Using a pre-trained audio encoding model, the proposed method encodes audio into a new token, which can be considered as an adaptation layer between the audio and text representations. Such a modeling paradigm requires a small number of trainable parameters, making the proposed approach appealing for lightweight optimization. Results suggest the proposed method is superior to the evaluated baseline methods, considering objective and subjective metrics.
    Code and samples are available at: \url{https://pages.cs.huji.ac.il/adiyoss-lab/AudioToken}.
\end{abstract}


\section{Introduction}
Neural generative models have changed the way we consume digital content. From generating high-quality images~\cite{sahariaphotorealistic, gafni2022make, chang2023muse}, though coherence of long spans of text~\cite{touvron2023llama, scao2022bloom,brown2020language}, up to speech and audio~\cite{kreuk2022audiogen,kreuk2022audio,lakhotia2021generative,wang2023neural}. In recent years, diffusion-based generative models have emerged as the preferred approach, showing promising results on various tasks~\cite{cao2022survey}.

\begin{figure}[t!]
    \begin{centering}
         \begin{subfigure}[b]{0.18\textwidth}
             \centering
             \includegraphics[width=\textwidth]{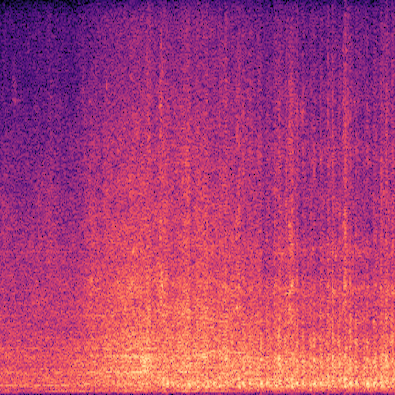}
         \end{subfigure}  ~~
         \begin{subfigure}[b]{0.18\textwidth}
             \centering
             \includegraphics[width=\textwidth]{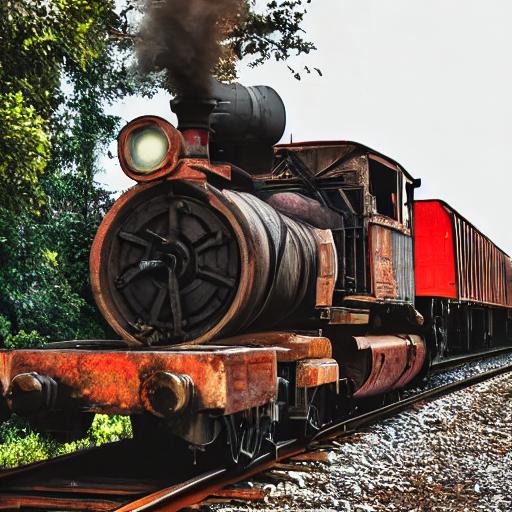}
        \end{subfigure}     
         \begin{subfigure}[b]{0.18\textwidth}
             \centering
                      \includegraphics[width=\textwidth]{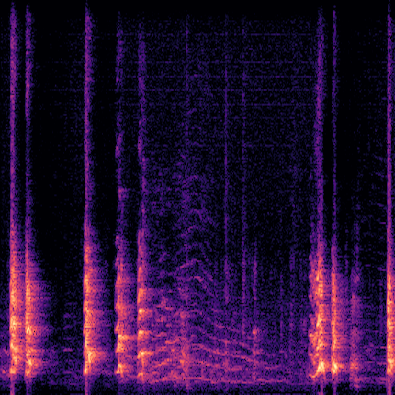}
         \end{subfigure}  ~~
         \begin{subfigure}[b]{0.18\textwidth}
             \centering
             \includegraphics[width=\textwidth]{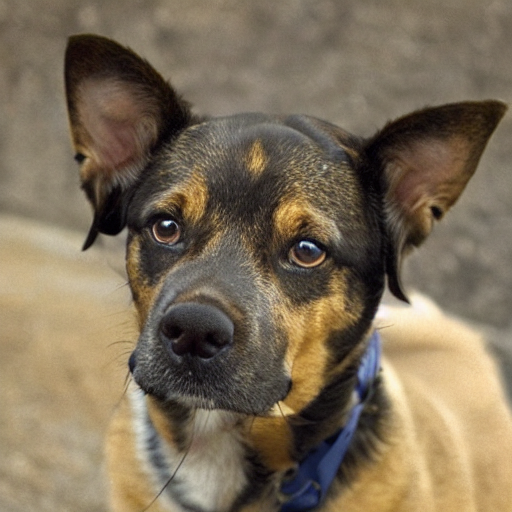}
             \end{subfigure}
         \caption{Generated images (right) and input spectrograms (left) from the proposed method. The model gets as input an audio recording, extracts a representation, and projects into a textual latent space which will be fed into a pre-trained text-conditioned diffusion generative model.}
         \vspace{-0.3cm}
         \label{fig:qual_fig}
    \end{centering}
\end{figure}

During the diffusion process, the model learns to map a pre-defined noise distribution to the target data distribution. In every step of the diffusion process, the model learns to predict the noise at a given step to finally generate the signal from the target distribution~\cite{ho2020denoising, dhariwal2021diffusion, nichol2021improved}. Diffusion models operate on different forms of data representations, e.g., raw input~\cite{kong2020diffwave,ho2020denoising}, latent representations~\cite{rombach2022high}, etc. 

\begin{figure*}[t!]
  \centering
  \includegraphics[width=0.8\linewidth]{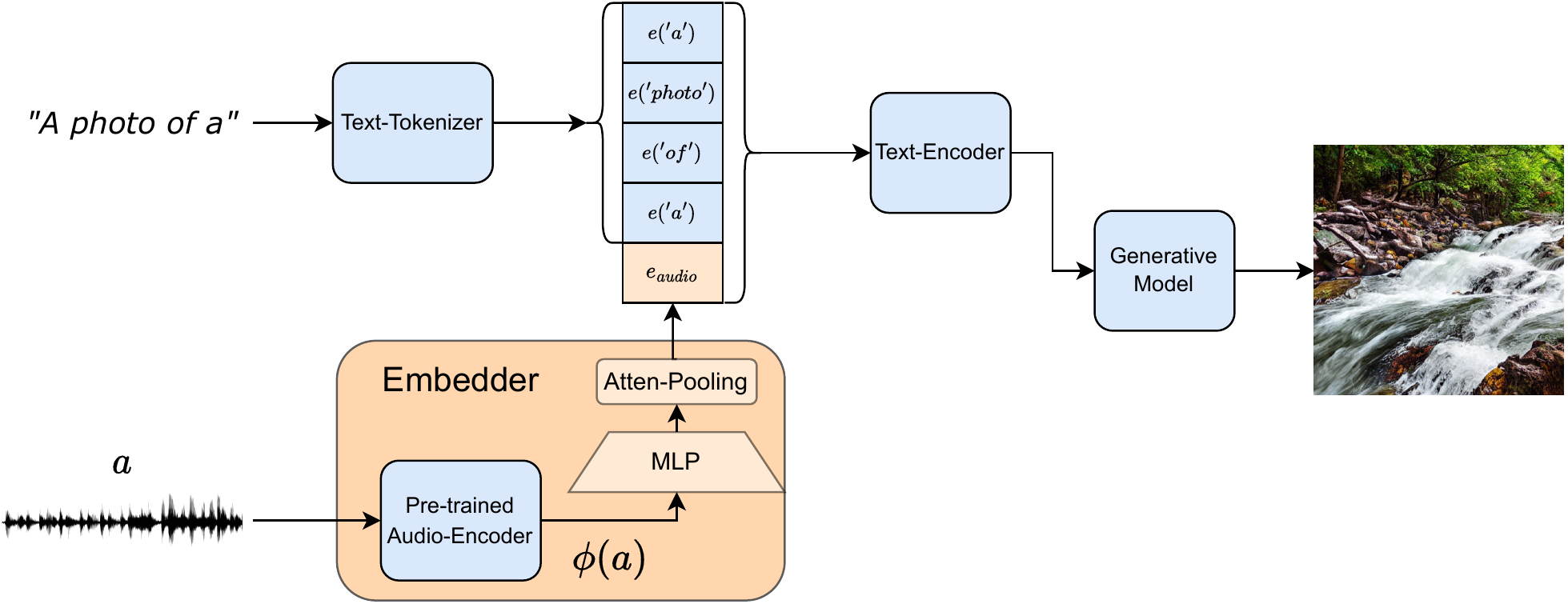}
  \caption{Architecture overview: We forward an audio recording through a pre-trained audio encoder and then through an Embedder network. A pre-trained text encoder extracts tokens created by a tokenizer and the audio token. Finally, the generative model is fed with the concatenated tensor of representations. It is important to note that only the Embedder parameters are trained during this process.}\vspace{-15pt}
  \label{fig:arch}
\end{figure*}

When considering controllable generative models, the common practice these days is to condition the generation on a textual description of the input data; this is especially noticeable in image generation~\cite{sahariaphotorealistic,ramesh2022hierarchical, nichol2022glide}.  Recently, several methods proposed using different modalities to condition the generative process such as image-to-audio~\cite{sheffer2022hear,SpecVQGAN_Iashin_2021}, image-to-speech~\cite{gao2021visualvoice,hsu2022revise}, image-to-text~\cite{tewel2022zerocap, li2023blip}, or audio-to-audio~\cite{kreuk2022textless, maimon2022speaking}. However, such research direction is less explored by the community.

In this work, we focus on the task of audio-to-image generation. Given an audio sample contains an arbitrary sound, we aim to generate a high-quality image representing the acoustic scene. We propose leveraging a pre-trained text-to-image generation model together with a pre-trained audio representation model to learn an adaptation layer mapping between their outputs and inputs.  Specifically, inspired by recent work on textual-inversions~\cite{gal2022image}, we propose to learn a dedicated \emph{audio-token} that maps the audio representations into an embedding vector. Such a vector is then forwarded into the network as a continuous representation, reflecting a new word embedding.

Several methods for generating audio from image inputs were proposed in prior work. The authors in~\cite{zelaszczyk2022audio,wan2019towards} proposed to generate images based on audio recordings using a Generative Adversarial Network (GAN) based method. Unlike the proposed method, in~\cite{zelaszczyk2022audio}, the authors present results for generating MNIST digits only and did not generalize to general audio sounds. In~\cite{wan2019towards}, the authors did generate images from general audio. However, this turned into low-quality images. The most relevant related work to ours is Wav2Clip~\cite{wu2022wav2clip}, in which the authors first learn a Contrastive Language-Image Pre-Training (CLIP)~\cite{radford2021learning} like a model for audio-image pairs. Then, later on, such representation can be used to generate images using VQ-GAN~\cite{esser2021taming} under the VQ-GAN CLIP~\cite{crowson2022vqgan} framework.

\emph{Why use audio signals as a conditioning to image generation rather than text?} Although text-based generative models can generate impressive images, textual descriptions are not naturally paired with the image, i.e., textual descriptions are often added manually. On the other hand, when considering videos, audio, and images capture and represent the same scene, hence are naturally paired. Moreover, audio signals can represent complex scenes and objects such as different types of the same instrument (e.g., classic guitar, electric guitar, etc.), or different scenes of the same object (e.g., classic guitar recorded in studio vs. live show). Annotating such fine-grained details of the different objects is labor-intensive, hence hard to scale.

In summary, our contributions are: We propose a novel method \audiotok~for audio-to-image generation by leveraging a pre-trained text-to-image diffusion model together with a pre-trained audio encoder; We propose a set of evaluation metrics specifically dedicated for the task of an audio-to-image generation. Through extensive experiments, we show that our method is able to generate high-quality and diverse set of images based on audio-scenes.  

\section{Adaptation of text-conditioned models}
Diffusion models are a family of models that are prone to learn the underlying probabilistic model of the data distribution $p(x)$. This is done by learning the reverse Markov process of length $T$. Given a timestamp $t \in [0, 1]$, the denoising function $\epsilon_{\theta}: \R^d \rightarrow \R^d$ learns to predict a clean version of the perturbed $x_t$ from the training distribution $S=\{x_1, .., x_m\}$:
\[
    \mathcal{L}_{\text{DM}} \triangleq \E_{x\sim S, t\sim U(0, 1),\epsilon\sim\mathcal{N}(0, I)}\left[\normx{\epsilon - \epsilon_{\theta}(x_t, t)}_2^2\right].
\]
Empirical results showed that learning diffusion models on top of latent spaces of autoencoders can produce results in a higher quality than those that are trained on the raw input~\cite{rombach2022high}. Intuitively, this process can be done on a latent representation of encoder-decoder architecture. Latent diffusion operates on top of a representation given by an encoder $f$:
\begin{equation}
    \mathcal{L}_{\text{LDM}} \triangleq \E_{x\sim S, t\sim U(0, 1),\epsilon\sim\mathcal{N}(0, I)}\left[\normx{\epsilon - \epsilon_{\theta}(f(x_t), t)}_2^2\right].
    \label{eq:ldm}
\end{equation}
The output of the diffusion can later be forwarded through the decoder to obtain the raw result (e.g., audio, image, text).

An important component of modern generative models is conditioning. This allows the generative process to be conditioned on a given input, i.e., modeling $p(x|y)$ where $y$ is a data entry. For example, in a text-based visual generation, the generative process is conditioned on text. There are many types of conditioning, such as text, time, style, etc.~\cite{rombach2022high}. Usually, the conditioning component is done by an injection of a condition representation from an encoder $\tau$ to the attention mechanism of $\epsilon_{\theta}$. Conditioning the diffusion process yields the following diffusion process, $\mathcal{L}_{\text{CLDM}} \triangleq$
\[
    \E_{(x, y)\sim S, t\sim U(0, 1),\epsilon\sim\mathcal{N}(0, I)}\left[\normx{\epsilon - \epsilon_{\theta}(f(x_t), t, \tau(y))}_2^2\right].
\]
In the following, we propose a method that leverages a conditional generative model to produce high-quality and diverse images that are based on audio-scenes. 

\subsection{\audiotok}

Audio signals contain information that can help us imagine the scene that produced them. This makes it tempting to use a generative model that is conditioned on audio recordings to generate a scene. However, models that generate high-quality images commonly rely on large-scale text-image pairs to generate images using text. We thus propose a method named \audiotok~that effectively projects audio signals into a textual space, enabling us to leverage existing text-conditioned models to generate images based on audio-based tokens.

Our objective is to investigate the feasibility of directly encoding any audio signals into a dedicated representation that will fit as an additional token for text-conditioning. By doing so, we can leverage existing pre-trained models and not  learn a new generative model with audio-visual pairs. Furthermore, we are not required to learn a new token for each individual class of audio or type of scene (as opposed to textual inversion-based methods~\cite{gal2022image}). Instead, we develop an audio-to-image generator capable of handling a wide range of diverse concepts.

The input to our method is a  short video input $(i, a)$, where $i$ represents a frame from the video and $a$ represents its corresponding audio recording. We are aiming to create a generative process that is audio-conditioned, i.e., $p(i|a)$. To achieve this, we utilize a text-conditioned generative model. Thus, we need to associate the audio signal $a$ with the text conditioning. 

The process begins with a transformer model that encodes the initial prompt ``A photo of a'' into a representation $e_{\text{text}}\in\mathbb{R}^{4\times d_a}$, where $d_a$ is the embedding dimension of the text input. Afterward, we concatenate to $e_{\text{text}}$, an extra latent representation of the audio signal, denoted as $e_{\text{audio}}\in\mathbb{R}^{d_a}$. We utilize an Embedder, which is composed of a pre-trained audio encoding network and a small projection network. This results in:
\[
    e_{\text{audio}} = \operatorname{Embedder}(a).
\]
Next, we describe the Embedder network and the optimization process of our method.

\noindent \textbf{Audio encoding:} The Embedder leverages a pre-trained audio classification network $\phi$ to represent the audio. The discriminative network's last layer is typically used for classification, and thus it tends to diminish important audio information which is irrelevant to the discriminative task. Thus, we take a concatenation of earlier layers and the last hidden layer (specifically selecting the fourth, eighth, and twelfth layers out of a total of twelve). This results in a temporal  embedding of the audio $\phi(a)\in\mathbb{R}^{\hat d\times n_a}$, where $n_a$ is the temporal audio dimension. Then, to learn a projection into the textual embedding space, we forward $\phi(a)$ in two linear layers with a GELU function between them:
\[
    \bar e_{\text{audio}} = W_2 \sigma (W_1 \phi(a)),
\]
where $W_1\in\mathbb{R}^{\hat d\times \hat  d}, W_1\in\mathbb{R}^{\hat  d\times  d_{\text{audio}}}$, and $\sigma$ is a GELU non-linearity~\cite{hendrycks2016gaussian}. Finally, we apply an attentive pooling layer~\cite{schwartz2019factor}, reducing the temporal dimension of the audio signal, i.e.,
\[
e_{\text{audio}} = \operatorname{Atten-Pooling}(\bar e_{\text{audio}}).
\]

\begin{figure}[t!]
     \centering
     \begin{subfigure}[b]{0.15\textwidth}
     \centering
     \includegraphics[width=\textwidth]{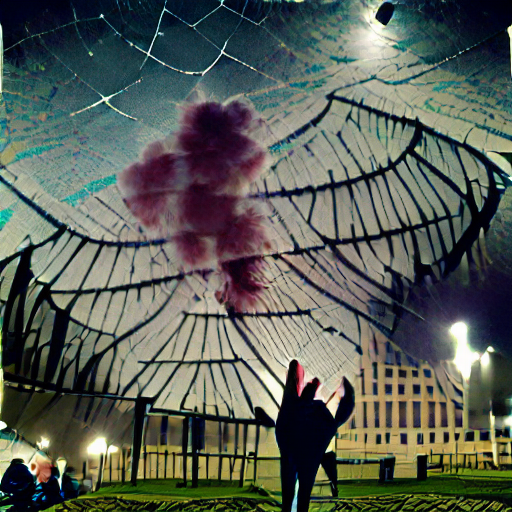}
     \end{subfigure}
     \begin{subfigure}[b]{0.15\textwidth}
     \centering
     \includegraphics[width=\textwidth]{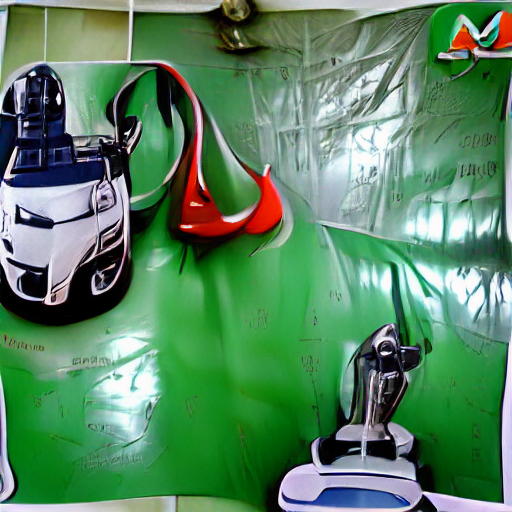}
     \end{subfigure}
     \begin{subfigure}[b]{0.15\textwidth}
     \centering
     \includegraphics[width=\textwidth]{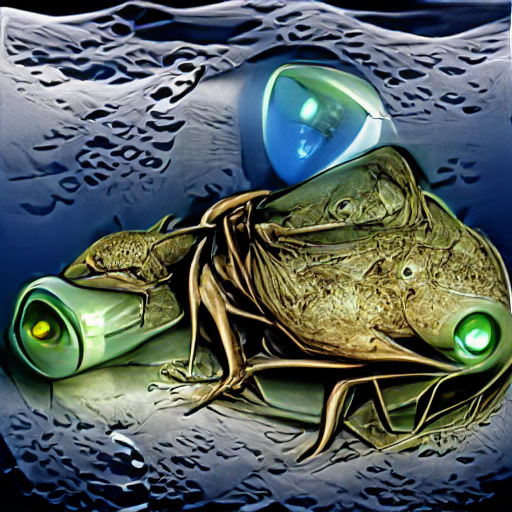}
     \end{subfigure}
    \\
         \centering
     \begin{subfigure}[b]{0.15\textwidth}
     \centering
     \includegraphics[width=\textwidth]{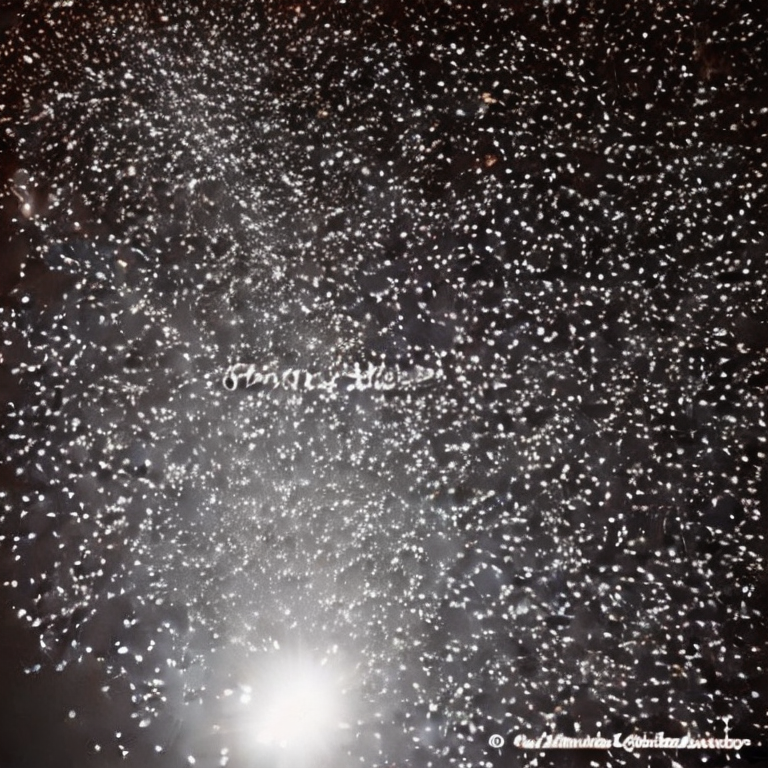}
     \end{subfigure}
     \begin{subfigure}[b]{0.15\textwidth}
     \centering
     \includegraphics[width=\textwidth]{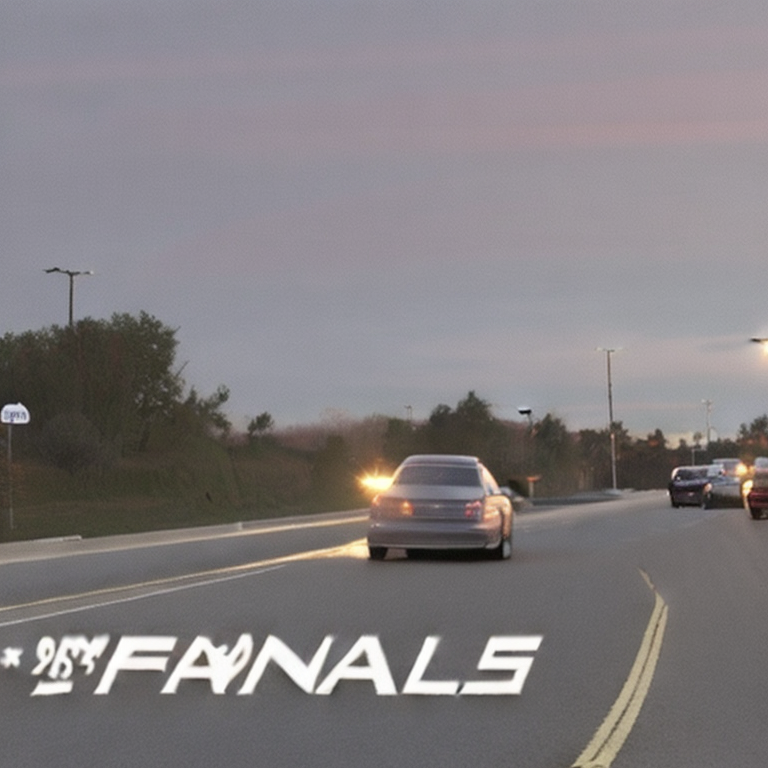}
     \end{subfigure}
     \begin{subfigure}[b]{0.15\textwidth}
     \centering
     \includegraphics[width=\textwidth]{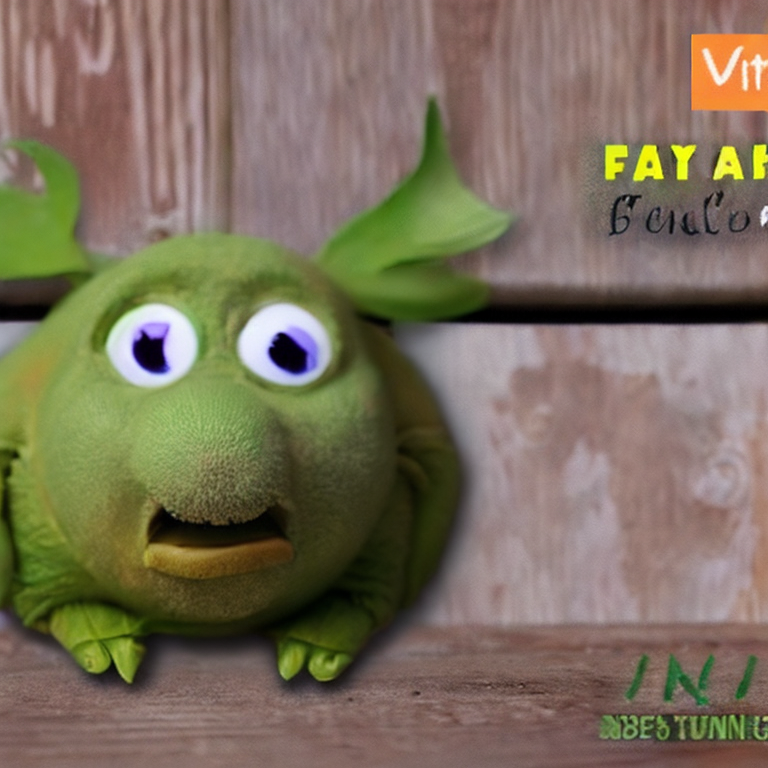}
     \end{subfigure}
    \\
    \begin{subfigure}[b]{0.15\textwidth}
     \centering
     \includegraphics[width=\textwidth]{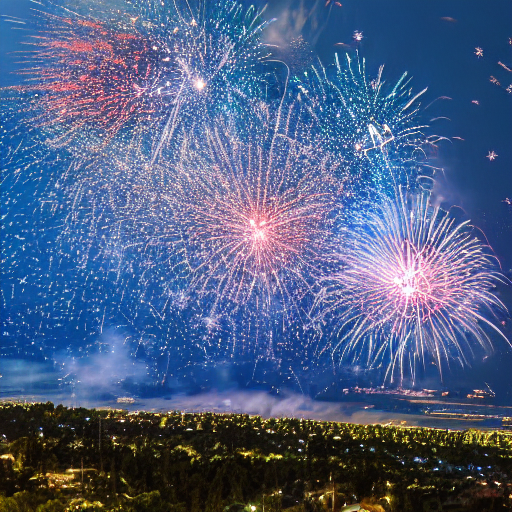}
     \end{subfigure}
     \begin{subfigure}[b]{0.15\textwidth}
     \centering
     \includegraphics[width=\textwidth]{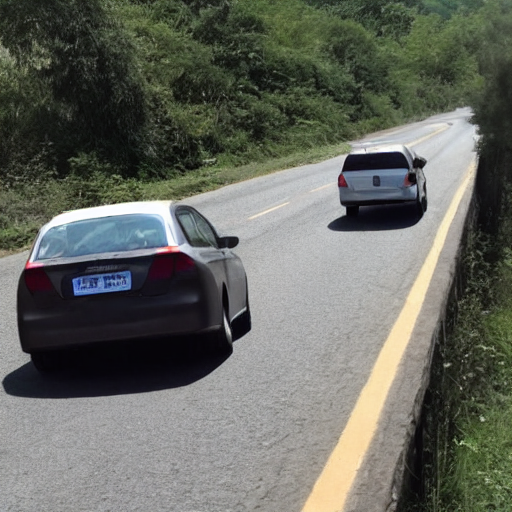}
     \end{subfigure}
     \begin{subfigure}[b]{0.15\textwidth}     
     \centering
     \includegraphics[width=\textwidth]{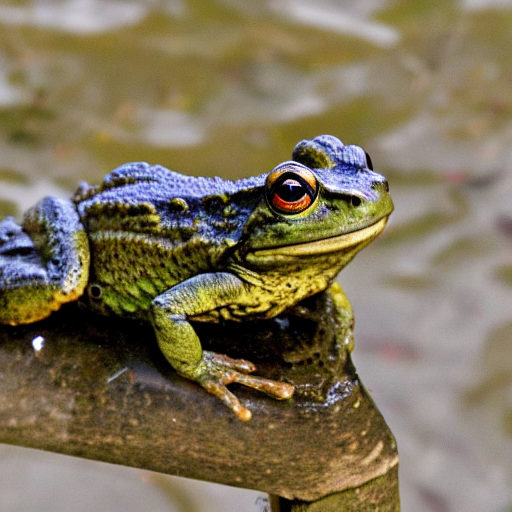}
    \end{subfigure}
     \\
     \begin{subfigure}[b]{0.15\textwidth}
     \centering
     \includegraphics[width=\textwidth]{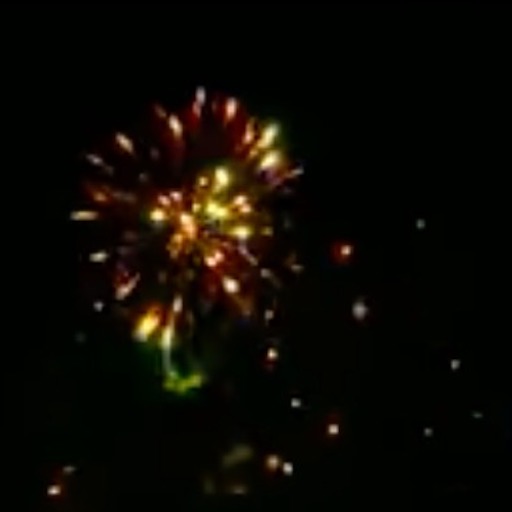}
     \end{subfigure}
     \begin{subfigure}[b]{0.15\textwidth}
     \centering
     \includegraphics[width=\textwidth]{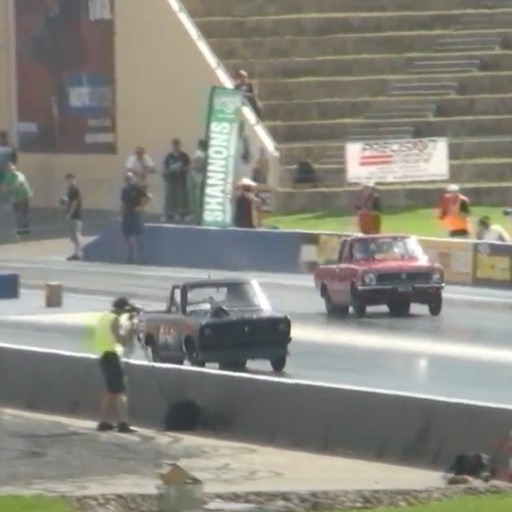}
     \end{subfigure}
     \begin{subfigure}[b]{0.15\textwidth}
     \centering
     \includegraphics[width=\textwidth]{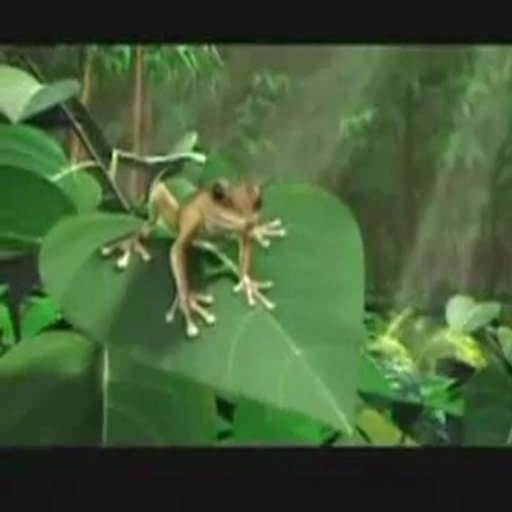}
     \end{subfigure}
     \caption{Qualitative results for Wav2Clip~(first row), ImageBind~(second row), \audiotok~(third row), and the original reference images (last row).}\vspace{-15pt}
     \label{fig:pull_fig}
\end{figure}

\noindent \textbf{Optimization:} During training, we update only the weights of the linear and attentive pooling layers within the Embedder network during the optimization process. The pre-trained audio network and the generative network remain frozen. We adopt the loss function employed by the original model $\mathcal{L}_{\text{LDM}}$ (Equation~\ref{eq:ldm}), maintaining consistency in the training scheme. Furthermore, we introduce an additional loss function that complements the original one, which involves encoding the label of the video, denoted by $l\in\mathbb{R}^{{n_l}\times d_{a}}$, where $n_l$ represents the label's length (e.g., the size of the `acoustic guitar' label is two). The label is encoded using the generative model's textual encoder, and then the spatial dimension is reduced using average pooling, i.e., $\hat l= \operatorname{Avg-Pooling}(l)$. The classification loss is defined as follows:
\[
    \mathcal{L}_{\text{CL}} = \left( 1 - \frac{\langle e_{\text{audio}}, \hat{l}\rangle}{\normx{e_{\text{audio}}}\normx{\hat l}} \right)^2.
\]
Intuitively, this term ensures that the audio embedding remains close to the video's concept, facilitating faster and more stable convergence. Finally, we also add an $\ell_1$ regularization to the encoded audio token, which encourages the audio token to be more evenly distributed. The overall loss that is optimized for \audiotok~is given by 
\[
    \mathcal{L} = \mathcal{L}_{\text{LDM}} + \lambda_{\ell_1}\normx{e_\text{audio}}_1.
\]
The overall loss that is optimized for \audiotok~with classification loss is given by 
\[
    \mathcal{L} = \mathcal{L}_{\text{LDM}} + \lambda_{\ell_1}\normx{e_\text{audio}}_1 +  \lambda_{\text{CL}}\mathcal{L}_{\text{CL}}.
\]

\subsection{Evaluation functions}\label{sec:evaluation_framework}

The evaluation of a visual generation from audio-scene is yet opened. Such evaluation setup is challenging since a well-performed model is excepted to generate images that will (i) capture the most prominent object in the audio recording; (ii) be semantically correlate with the input audio; and (iii) be semantically similar to the ``ground truth'' / target image. Lastly, we require evaluating the general quality of the generated image. To mitigate that, we propose to use the following evaluation functions. 

\noindent{\bf Audio-Image Similarity (AIS)} ideally measures the similarity between the semantic input audio and generated image features. We employ the Wav2CLIP model~\cite{wu2022wav2clip}. The Wav2CLIP model enables to measure of the similarity between representations of an audio and image pair. This allows us to quantify to which extent the generated image describes the audio.  Quantifying only the correlation score is not telling the whole story since the score scale may vary. Thus, it is unclear what is a good score. Instead, we compare the similarity between a generated image  and its input audio and the similarity between the generated image and arbitrary audio from the data. The AIS score is then averaged over all data entries in the validation set.

\noindent{\bf Image-Image Similarity (IIS)} measures the semantic similarity between the generated image and the ``ground truth'' one. This information is crucial since it allows quantifying the semantic similarity to a ``ground truth'' scene.

We employ the same reference-based method as in the AIS metric. Thus, we measure the CLIP~\cite{radford2021learning} score between the (i) generated image and its ``ground truth'' and (ii) generated image and an arbitrary image from the data. The IIS score is then averaged over all data entries in the validation set.

\noindent{\bf Audio-Image Content (AIC).} To account for the image content, we measure the level of agreement between the predicted class of an image classifier and the ground-truth audio label. However, since there might not be a complete correlation between the image classifier classes and the audio labels, an additional CLIP-based score is employed to determine agreement. If the CLIP-based matching score exceeds a threshold of 0.75, the image and audio class are considered in agreement. 

\noindent{\bf Fr\'echet Inception Distance (FID).} In order to evaluate the quality of the generated images, we adopt the standard FID score~\cite{heusel2017gans}. Such reference-free metric compares the distribution of the generated images against the original images using an internal representation obtained from a pre-trained model. In this work, we use the Inception model.

\noindent{\bf Human Evaluation.} Lastly, we run a subjective test to evaluate the adherence of the generated images to their labels. For each method, annotators were shown a generated image and asked to rate its relevance to a given label on a scale of 1-5. 

\begin{table}[t!]
\centering
\caption{We report AIC, FID, AIS, and IIS for \audiotok~(with and without Classification Loss (CL)), together with Wav2Clip. For reference, we additionally report results for the original images (reference) and images generated by Stable Diffusion (SD) with text labels.}
\vspace{-0.1cm}
\label{table:ued}
\resizebox{0.95\linewidth}{!}{
\begin{tabular}{lccccc}    
    \toprule
    \multirow{2}{*}{Method}& % \multirow{2}{*}{Labels}& 
    \multicolumn{4}{c}{Metric} \\
    \cmidrule(lr){2-5}
      & AIC $\uparrow$ & FID $\downarrow$ & AIS $\uparrow$ & IIS $\uparrow$\\
    \midrule
    Reference &54.66&-& - & -\\
    SD (Text) &{ 71.28} & {52.85}& - & -\\
    \midrule
    % Wav2Clip~\cite{wu2022wav2clip}& 16 &37.56 & 103.32 &  49.34& 53.87\\
    % \audiotok~w.o CL &16&\textbf{65.32} & \textbf{ 68.76}& \textbf{68.21}& \textbf{66.98}\\
    % \midrule
    Wav2Clip~\cite{wu2022wav2clip}&  29.32         & 99.89         & 47.76      & 51.11\\
    ImageBind~\cite{girdhar2023imagebind}&  39.15         & 67.42         & 67.48      & 75.50\\
    \audiotok~with CL & \textbf{48.01} & 66.08 & 62.28  & 76.40\\
    \audiotok & 45.48 & \textbf{56.65} & \textbf{68.23}  & \textbf{76.66} \\
    \bottomrule
\end{tabular}}
\vspace{-0.4cm}
\end{table}

\section{Results}
In the following, we study our method from objective and subjective points of view. We begin by describing details regarding the experimental setup. Then, we report results for our method and baselines using the evaluation framework proposed in Section~\ref{sec:evaluation_framework}. We show that our method outperforms the current baselines. We finally subjectively evaluate and find that annotators agree that our method describes the audio the best.

\noindent {\bf Baselines.} Wav2Clip~\cite{wu2022wav2clip} employs a CLIP-based loss for audio-text pairs. Then, they use this representation to generate an image from a text that is highly correlated with the audio using VQ-GAN~\cite{esser2021taming}. 
ImageBind~\cite{girdhar2023imagebind} combines information from six different modalities (text, image/video, audio, depth, thermal, and inertial measurement units (IMU)) into a single representation space. We used ImageBind's unified latent with stable-diffusion-2-1-unclip~\footnote{\url{https://github.com/Zeqiang-Lai/Anything2Image}} to generate images from audio samples.

\noindent {\bf Data.} We use the VGGSound~\cite{chen2020vggsound} dataset, which is derived from a collection of YouTube videos with corresponding audio-visual data. The dataset contains $200,000$, each in the length of ten seconds. The dataset is also annotated with $309$ classes. 

\noindent {\bf Hyperparameters.} The Embedder network comprises 3 layers, with attention pooling applied to a sequence of 248. For the generative model, we use Stable Diffusion~\cite{rombach2022high}. This results in an $8,853,507$ parameters model. During training,  we randomly crop five-second audio clips and select the frame with the highest CLIP score corresponding to the VGGSound label. We also filter out frames with inconsistent classifications from both the image and audio classifiers. We train the model for $60,000$ steps with a learning rate of 8e-5 and batch size of $8$ on Nvidia A6000 GPU.

\noindent\textbf{Objective evaluation.}
We start by comparing the proposed method, with and without the Classification Loss (CL), against Wav2Clip and ImageBind, considering FID, AIS, AIC, and IIS. For reference, we additionally include a topline of results of generating images directly from textual description (text labels) using Stable Diffusion (SD). Results are reported in Table~\ref{table:ued}.

Results suggest that $\audiotok$ is superior to Wav2Clip and ImageBind, considering all evaluation metrics. Interestingly, $\audiotok$ also performs better when considering the AIS metric, which leverages the Wav2Clip and ImageBind models to obtain the similarity score. This result demonstrates accurate audio detail identification (e.g., distinguishing various guitars) and considers multiple entities (e.g., multiple flying planes or a single plane). As expected, using textual labels reaches a higher accuracy and pushes the model toward learning representation which is more discriminative but less correlated with the target video.  Generated images from all methods can be seen in Figure~\ref{fig:pull_fig}.
\begin{figure}[t!]
     \centering
     \begin{subfigure}[b]{0.11\textwidth}
     \centering
     \includegraphics[width=\textwidth]{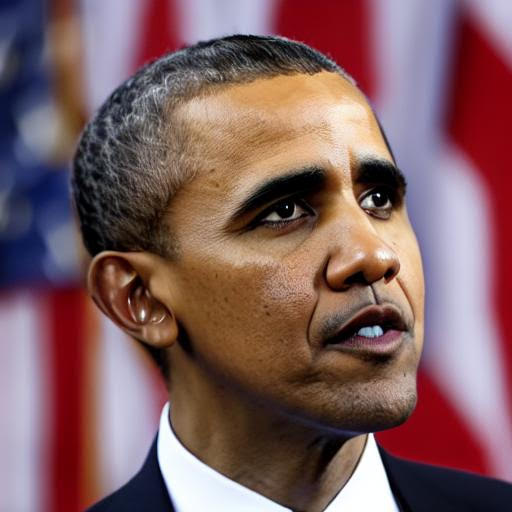}
     \end{subfigure}
     \begin{subfigure}[b]{0.11\textwidth}
     \centering
     \includegraphics[width=\textwidth]{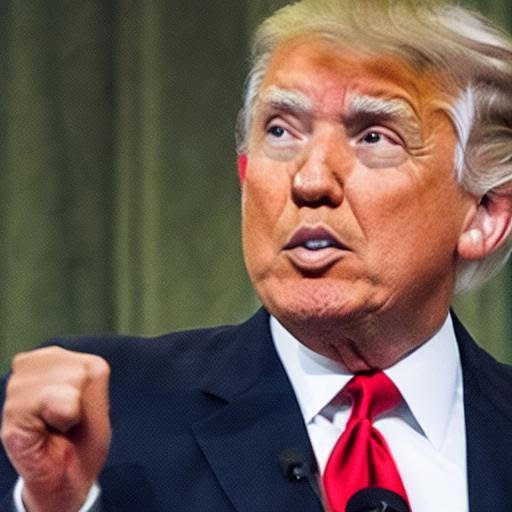}
     \end{subfigure}
     \begin{subfigure}[b]{0.11\textwidth}     
     \centering
     \includegraphics[width=\textwidth]{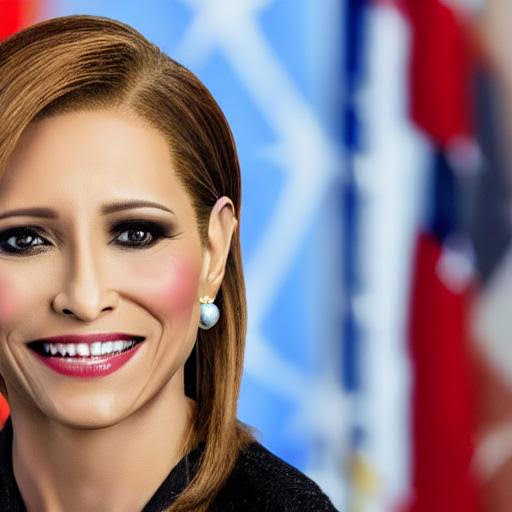} 
     \end{subfigure}
    \begin{subfigure}[b]{0.11\textwidth}
     \centering
     \includegraphics[width=\textwidth]{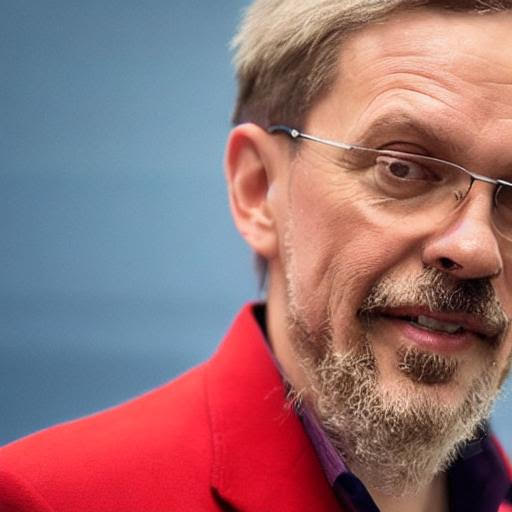}
     \end{subfigure}
     \\
     \begin{subfigure}[b]{0.11\textwidth}
     \centering
     \includegraphics[width=\textwidth]{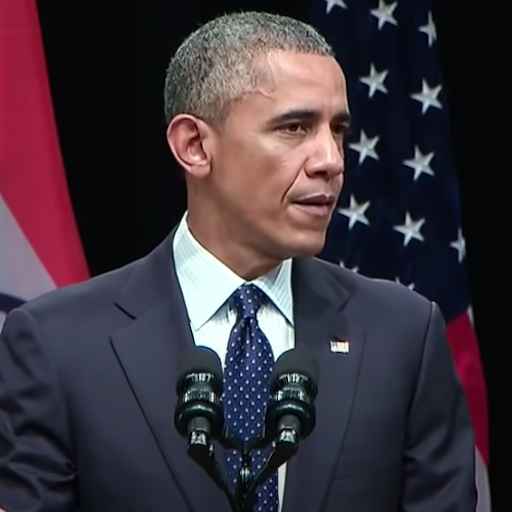}
     \end{subfigure}
     \begin{subfigure}[b]{0.11\textwidth}
     \centering
     \includegraphics[width=\textwidth]{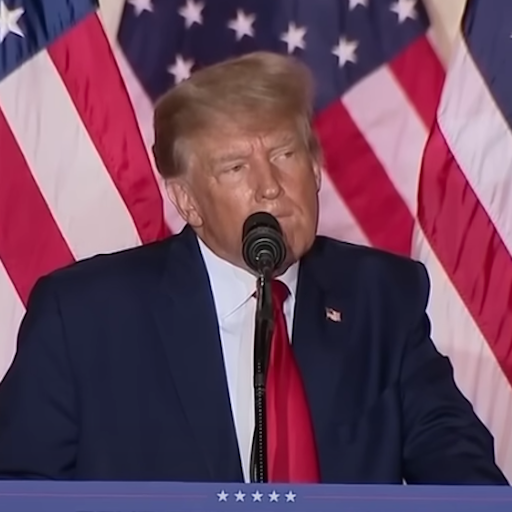}
     \end{subfigure}
     \begin{subfigure}[b]{0.11\textwidth}
     \centering
     \includegraphics[width=\textwidth]{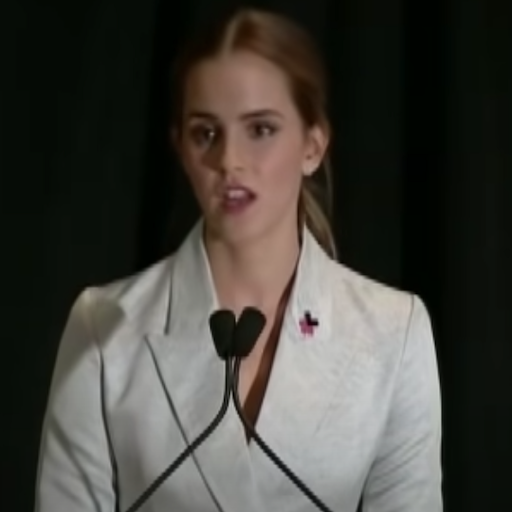}
     \end{subfigure}
          \begin{subfigure}[b]{0.11\textwidth}
     \centering
     \includegraphics[width=\textwidth]{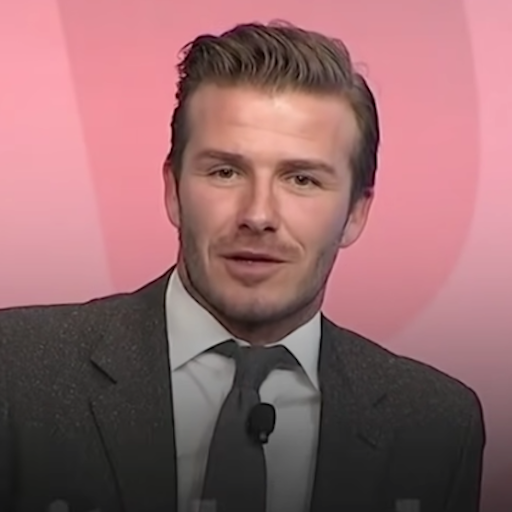}
     \end{subfigure}
     \caption{Qualitative results of speaker generation for \audiotok~(first row), and reference images (second row).} 
     \vspace{-15pt}
     \label{fig:celeb}
\end{figure}

\noindent\textbf{Subjective evaluation.}
We compare \audiotok~ against Wav2Clip, and SD using textual descriptions. We randomly sample 15 images from the test set and ask human annotators to rank their relevance to their textual labels on a scale between 1 and 5. We enforce at least 17 annotations for each of the evaluated images and compute the mean score together with its standard deviations.  \audiotok~outperforms Wav2Clip ({4.07 $\pm$ 0.83} vs. {1.85 $\pm$ 0.46}). When considering comparison to SD using text labels, \audiotok~ is reaches comparable performance and yields slightly worse subjective scores ({4.07 $\pm$ 0.83} vs. {4.58 $\pm$ 0.60}). These findings are especially encouraging, as these suggest users found the images generated by \audiotok to capture the main objects in the audio scene similarly to using textual labels, which serves as a topline.

% \vspace{0.1cm}
\noindent\textbf{Speaker image generation.}
We investigate its potential to create visuals of various speakers. We gathered samples from two 30-minute videos per person that showcased Barack Obama, Donald Trump, Emma Watson, and David Beckham to achieve this goal and extracted the audio representation from X-Vector~\cite{snyder2018x}. Our results in Fig.~\ref{fig:celeb} indicate that our approach accurately represents Barack Obama and Donald Trump. We postulate that this could be due to their distinct voices. However, with Emma Watson and David Beckham, the method mainly captures their gender.

\vspace{-0.2cm}
\section{Conclusions}
In this paper, we present a method for leveraging text-conditioned generative models for audio-based conditioning. Our method produces high-quality images which describe a scene from the audio recording. In addition, we propose a comprehensive evaluation framework that takes into account the semantics of the images generated. Our method presents a first step toward audio-conditioned image generation. The hidden information in the audio is richer than the observed one in the text. Hence, we think that this problem is interesting and should get more focus from the community.

\clearpage

\bibliographystyle{IEEEtran}
\bibliography{refs}

\end{document}